\documentstyle[preprint,aps,epsfig]{revtex}
\newcommand{\lsim}{\mathrel{\mathop{\kern 0pt \rlap
  {\raise.2ex\hbox{$<$}}}
  \lower.9ex\hbox{\kern-.190em $\sim$}}}
\newcommand{\gsim}{\mathrel{\mathop{\kern 0pt \rlap
  {\raise.2ex\hbox{$>$}}}
  \lower.9ex\hbox{\kern-.190em $\sim$}}}

\newcommand{\tchi}     {\tilde{\chi}}
\newcommand{\stau}     {{\tilde{\tau}}}

\newcommand{\gm}     {\gamma}
\newcommand{\Gm}     {\Gamma}
\newcommand{\lm}     {\lambda}
\newcommand{\sg}     {\sigma}

\newcommand{\Lg}     {{\mathcal L}}
\newcommand{\R}     {{\mathcal R}}
\newcommand{\beq}     {\begin{equation}}
\newcommand{\eeq}     {\end{equation}}
\newcommand{\bea}     {\begin{eqnarray}}
\newcommand{\eea}     {\end{eqnarray}}

\begin{document}
\draft
\preprint{
\vbox{\hbox{hep-ph/0107218}
      \hbox{KIAS P01032}
      \hbox{AS-ITP-2001-013}
      }
}

\title{
Stau LSP and comparison with $H^\pm$ phenomenology
}

\author{
A.G. Akeroyd$^{\:a}$\footnote{E--mail: akeroyd@kias.re.kr},
Chun Liu$^{\:b}$\footnote{E--mail: liuc@itp.ac.cn},
and Jeonghyeon Song$^{\:a}$\footnote{E--mail: jhsong@kias.re.kr}
}

\vspace{1.5cm}

\address{
$^a$Korea Institute for Advanced Study, 207-43 Cheongryangri-dong,
Dongdaemun-gu,\\
Seoul 130-012, Korea\\
$^b$Institute of Theoretical Physics, Chinese Academy of Sciences,\\
P.O. Box 2735, Beijing 100080, China
}

\maketitle

\thispagestyle{empty}

\setcounter{page}{1}

\begin{abstract}
In supersymmetric models with explicit breaking of $R$-parity
the lightest supersymmetric
particle (LSP) may be the lightest stau, $\stau_1$.
Such a scenario
would provide a clear sign of $R$-parity violating SUSY,
although
its phenomenology may resemble
that of a charged Higgs boson, $H^\pm$.
We discuss various ways of distinguishing a LSP $\stau_1$
from $H^\pm$ at future colliders, and address the
case of $\stau_1$ mimicking the signal for $H^\pm$.
As an example we suggest that the recent L3 signal for
$H^+H^-\to qq'qq'$ and $H^+H^-\to qq'\tau\nu_{\tau}$ could be
more easily explained by a LSP $\stau_1$.
\end{abstract}


\newpage

\section{Introduction}
\label{introduction}

$R$-parity violating extensions of the
minimal supersymmetric standard model (MSSM) have received
much attention
since the observation of neutrino oscillations
\cite{Fukuda:1998mi}.
Small neutrino masses can be naturally generated
through trilinear and/or bilinear lepton number violating
interactions \cite{neutrino,Hirsch:2000ef}.
Moreover the particle content of the MSSM remains intact.
A clear signal of $R$-parity violation would be the single
production of supersymmetric particles,
and/or observation of a charged lightest supersymmetric
particle (LSP).
The latter situation is allowed in
a $R$-parity violating supersymmetric model
because the LSP is unstable.
A charged or colored $stable$ LSP
would conflict with cosmological observations
by forming readily detectable anomalous heavy isotopes \cite{Rich:1987jd}.
If the LSP is unstable then such cosmological constraints become irrelevant.
In this paper we focus on the
case of the LSP being the lightest scalar tau,
$\stau_1$.

In a general $R$-parity violating supersymmetric model,
the phenomenology of the stau has been known
to possess many similarities with
that of the charged Higgs boson $H^\pm$
{ \cite{Dreiner:1991pe,Borzumati:1999th}}.
For example,
at future $e^+e^-$ colliders both can be pair-produced by the
same mechanism $e^+e^-\to \stau_1^+\stau_1^-,H^+H^-$
with very similar rates, especially
if the $\stau_1$ is mainly left-handed. Therefore
distinguishing $\stau_1$ from $H^\pm$
is an issue of significance and importance at future colliders.
There are, in principle, at least
two ways in which $\stau$ and $H^\pm$ may differ phenomenologically;
the mass spectrum and the decay modes. Firstly in the MSSM,
the mass of $H^\pm$ ($M_{H^\pm}$) originates from the
supersymmetric and gauge--invariant superpotential, and
at tree--level $M_{H^\pm}$ is related to the pseudoscalar mass
$M_A$ and the $W$ boson mass $M_W$
by $M_{H^\pm}^2=M_A^2+M_W^2$.
Although this relationship may be relaxed in extensions of the MSSM, or
in other non--supersymmetric models with an extended Higgs sector,
the contribution of $H^\pm$ to the decay $b \to s \gm$
often imposes a strong lower bound on $M_{H^\pm}$.
In comparison, the presence of
a light $\stau_1$ is compatible with the experimental measurement
of $b \to s \gm$.
Secondly, decays of $H^\pm\to ff'$  are proportional
to the mass of the fermion and involve
the parameter $\tan\beta$ $(=v_2/v_1$).
Therefore for a given $M_{H^\pm}$ the branching ratios (BRs)
are calculable functions of $\tan\beta$.
In contrast, $\stau_1\to
ff'$ decays involve the arbitrary $R$-parity violating
couplings $\lambda$ and $\lambda'$.
Therefore in general there are many more decay possibilities for
$\stau_1\to ff'$ \cite{Dreiner:1991pe,Borzumati:1999th}.

Furthermore our assumption of a LSP $\stau_1$
can provide a unique phenomenology which includes
the possibility of $\stau_1$ closely mimicking
$H^{\pm}$. If $M_{H^\pm}\le m_t$ then
$H^\pm$ decays mainly into $cs$ and $\tau\nu$
and the misidentification can occur if the relevant
$\lm$ and $\lm'$ couplings are both non--zero.
This possibility has not yet been seriously considered
due to the usual assumption that one $R$-parity
violating operator is dominant at a time
while the others are negligible.
LEP has carried out searches for
$\stau_1$ as the LSP but the dominance of
one coupling is always assumed \cite{Abbiendi:2001is}.
The possibility of misidentifying $\stau_1$ and
$H^\pm$ has important implications for future colliders,
and merits further experimental and theoretical consideration.
Any signal in a given search for $H^{\pm}$ should be interpreted
with care in order to be sure
that the signal really is a $H^{\pm}$.
Observation of $H^{\pm}$ is expected
to provide a useful measurement of $\tan\beta$
since this parameter strongly determines its phenomenology.
Therefore, measurements of $\tan\beta$ from $H^\pm$-like signals
should await complementary confirmation.

An example of $\stau_1-H^\pm$ misidentification
affecting the interpretation of experimental signals
is the recently reported
L3 excess of $4.4\sigma$ in the channels
$H^+H^-\to qq'qq'$ and $H^+H^-\to qq'\tau\nu_{\tau}$
\cite{Garcia-Abia:2001eb}. The data is compatible with
$M_{H^\pm}=68\,$GeV,
BR$(H^\pm\to cs)\approx 90\%$ and BR$(H^\pm\to
\tau\nu_{\tau})\approx 10\%$.
As shall be discussed later,
many popular extensions of the SM have difficulties in
incorporating such light charged Higgs bosons.
Neither the MSSM, nor the next to MSSM (NMSSM),
nor two Higgs doublet models (2HDM)
can accommodate the above L3 results.
In contrast, a $\stau_1$ with mass around 68\,GeV
and with observed branching ratios can be obtained
within a reasonable SUSY parameter space.
Note that the LSP requirement for the stau is
crucial, since otherwise its dominant decay mode
would be into a tau lepton ($\tau$) and a LSP neutralino ($\chi^0$).
The latter decays would give rise to a signature incompatible with
the observed BRs (i.e., missing energy for $\chi^0$
if $R$-parity is conserved, or high multiplicity fermionic
events if $\chi^0$ decays).

This paper is organized as follows.
In Sec.~\ref{stau} we specify the conditions for
a LSP $\tilde \tau_1$ in the context of
a $R$-parity violating model.
In Sec.~\ref{L3} we briefly summarize the L3 results
and consider its interpretation as a $H^{\pm}$
in some popular supersymmetric and
non--supersymmetric models with two or more Higgs doublets.
It shall be shown that it is difficult to
accommodate the L3 excess except for
models with 3 or more doublets. In Sec.~\ref{stau-L3}
we offer an attractive explanation of the L3 data
in terms of the stau LSP.
In Sec.~\ref{diff}
we compare the phenomenology of $\tilde\tau_1$ and
$H^\pm$ at future colliders, and specify the conditions
where the phenomenology is very close
and where it can differ.
Finally, Sec.~\ref{conclusion} contains our conclusions.

\section{Stau as the LSP in the MSSM with explicit $\rlap/R_p$}
\label{stau}

Explicit $R$-parity violation
in the MSSM is generated by adding
all possible renormalizable
$L$-violating couplings to the superpotential
\cite{Dreiner:1997uz}:
\beq
W_{\rlap/R _p}
= \frac{1}{2} \lm_{ijk} L_i L_j \bar{E}_k
+\lm'_{ijk} L_i Q_j \bar{D}_k,
\eeq
where $i,j,k=1,2,3$ are generation indices,
$\lm_{ijk}=-\lm_{jik}$,
$L_i ~(Q_i)$
are the lepton (quark) $SU(2)$ doublet superfields,
and $\bar{E}_i ~ (\bar{D}_i)$
are the lepton (down-quark) $SU(2)$
singlet superfields.
The $B$-violating couplings $\lm'' \bar{U}\bar{D}\bar{D}$
are set to zero in order to forbid proton decay.
In addition, a bilinear term $\epsilon_iL_i H_2$ may be
added, which generates a tree--level mass for neutrinos
through their mixing with the neutralinos.
Such a bilinear term is also known to induce mixing between
$H^\pm$ and $\stau_i$ \cite{akediaz}. The smallness of the
neutrino mass indicated by the Super-Kamiokande data would
suggest a suppressed $\epsilon_i$\cite{Hirsch:2000ef}, and thus small
mixing between $H^\pm$ and $\stau_i$. If this mixing is unsuppressed
then a LSP $\stau_1$ in a purely bilinear $R$-parity violating model
would decay promptly via its $H^\pm$ component, with $H^\pm$ like BRs
\cite{akediaz}. For suppressed mixing $\stau_1$ would decay
with a long lifetime, again with $H^\pm$ like BRs.
We shall be working in the $R$-parity violating model
defined by the superpotential above, and we shall see that
the large range of values of the
$\lambda$ and $\lambda'$ couplings
enable the LSP $\stau_1$ to have a richer phenomenology than
that of the LSP $\stau_1$ in the purely bilinear case.

The $R$-parity violating Yukawa interactions
involving sleptons are given by,
in four-component Dirac notation,
\beq
\label{Lg}
\Lg =
\lm_{ijk}
\left[
\tilde{e}^j_L \bar{e}^k_R\nu^i_L
+
(\tilde{e}^k_R)^* (\bar{\nu}^i_L)^c e_L^j
- ( i\leftrightarrow j)
\right]- \lm'_{ijk}\tilde{e}^i_L
\bar{d}^{k}_R u^j_L +H.c.\,.
\eeq
Since the sleptons can decay into SM leptons and/or quarks,
the cosmological condition
for the LSP to be charge- and color-neutral
becomes inapplicable.
We restrict ourselves to phenomenological implications
of the case where the stau is the LSP.

The mass matrix squared for the
left- and right--handed stau's
(neglecting possible $CP$-violating phases) is given by
\begin{eqnarray}
{\cal M}^{\,2}_{\tilde{\tau}}
=\left(\begin{array}{cc}
          X_\tau
          & Z_\tau\\[2mm]
Z_\tau &       Y_\tau
\end{array}\right)\,,
\end{eqnarray}
where $X_\tau$, $Y_\tau$, and $Z_\tau$ are
\begin{eqnarray} \label{eq:xyztau}
X_\tau&=&m^2_{\tilde{\tau}_{_L}} +m^2_\tau
       +\frac{1}{2}(m^2_Z-2m^2_W)\cos 2\beta\,,\nonumber\\
Y_\tau&=& m^2_{\tilde{\tau}_{_R}} +m^2_\tau
       +(m^2_W-m^2_Z)\cos 2\beta\,, \nonumber\\
Z_\tau&=&m_\tau\,|\,A_\tau + \mu\, \tan\beta\,|\,.
\end{eqnarray}
Here  $m^2_{\tilde{\tau}_{_{L,R}}}$ are respectively
the left- and right-handed
soft-SUSY-breaking stau masses squared; $A_{\tau}$ is the soft
trilinear coupling for the $\stau$. Diagonalizing this
matrix leads to two mass eigenstates $\stau_1$ and $\stau_2$,
with $m_{\stau_2}\ge m_{\stau_1}$.

As one can see, the mass of $\stau_i$ depends on a combination
of $\mu$, $\tan\beta$ and soft SUSY breaking parameters,
all of which are very
weakly constrained by experiment.
This is in contrast to
$M_{H^\pm}$ in the MSSM, which is constrained by the sum rule
obtained from the scalar potential.
Hence a light $\stau_1$ is permitted if one of
$m^2_{\tilde{\tau}_{_L}}$ and $m^2_{\tilde{\tau}_{_R}}$ is
chosen to be suitably small. Therefore a LSP $\stau_1$
is certainly possible in the $R$-parity violating MSSM.
We note that models
which assume universality of scalar masses
at the GUT scale will not in general produce a LSP $\stau_1$.
Models with anomalous breaking of supersymmetry generally
require the sleptons to be lighter than the other
SUSY particles\cite{Feng:2000hg},
and thus a LSP stau may arise in such models provided
$R$-parity is also broken.

A comment on constraints from the decay $b\to s \gamma$
is in order here. It is known that
a light charged Higgs boson
($M_{H^\pm} < M_W$) can give an unacceptably large contribution
to the measured decay $b \to s \gm$, as shall be discussed in
the next section. However, $R$-parity violating supersymmetric models
have been shown to be weakly constrained by the $b \to s\gm$ decay
due to the large number of free parameters
coming from new (complex) Yukawa couplings\cite{Besmer:2001rj}.
In the scenario of the LSP stau, the dominant contribution
to $b \to s \gm$
from the $R$-parity violating Yukawa interactions
in Eq.~(\ref{Lg}) is mediated by $\stau_1$ and a top quark,
giving a contribution proportional to
$| \lm'_{333} \lm'_{332}|^2$.
The adjustment of these parameters is, in principle, always possible
to avoid the $b \to s \gm$ bounds.
Therefore a LSP $\stau_1$ is certainly a viable option
in the $R$-parity violating MSSM.

\section{L3 excess and $H^\pm$ interpretation}
\label{L3}

Based on the recent search
for pair--produced charged Higgs bosons
with data collected
at $200\, {\rm GeV} \le \sqrt s\le 209 \,{\rm GeV}$,
the L3 collaboration has reported
signals in the channels $H^+H^-\to c\bar{s} \,\bar{c} s$ and
$H^+H^-\to c\bar{s}\,\tau^-\bar{\nu}_{\tau}$ \cite{Garcia-Abia:2001eb}.
The data is compatible with a 4.4\,$\sigma$ fluctuation
in the background, and is best fitted by a $H^{\pm}$
with $M_{H^\pm}=68$\,GeV,
BR$(H^\pm\to cs)\approx 90\%$ and BR$(H^\pm\to
\tau\nu_{\tau})\approx 10\%$.
Although similar excesses have not been
observed by OPAL, DEPLHI, and ALEPH,
the full confirmation of the L3 results
still awaits future experiments.
There is a possibility that differences in
the search strategies among the four collaborations
may be a partial explanation of why the above three experiments
have not observed the L3 excess. In particular, the DELPHI
search utilizes $c$-tagging since $H^\pm\to cs$ is expected to be the
dominant quark decay channel. Note that the DELPHI search
would not be sensitive to
anomalous decay modes of the charged scalar, e.g.,
decays to light quark jets. The compatibility
of the four experiments is currently
being investigated by the LEP working groups \cite{Group:2001xy}.

Discovery of a $H^{\pm}$ would be immediate evidence of physics beyond
the minimal SM, since the latter predicts the existence
of only a single neutral Higgs boson.
In various models the Higgs sector is extended to include
two or more Higgs doublets,
leading to a physical Higgs spectrum with charged Higgs bosons.
The MSSM requires two Higgs doublets, and the supersymmetric
structure of the theory imposes constraints on the Higgs potential.
This constrained tree--level Higgs potential ensures the
following sum rule
\cite{MSSM,HHG}:
\begin{equation}
M_{H^\pm}^2=M_A^2+M_W^2.   \label{sum}
\end{equation}
Equation (\ref{sum}) is only significantly affected
by one--loop corrections in the parameter space
of very low $\tan\beta$ \cite{Diaz:1992ki},
which is now experimentally excluded.
The current lower bound from LEP $M_A\ge 90$\,GeV
implies $M_{H^\pm}\ge 110$\,GeV,
taking $H^\pm$ out of the discovery reach of LEP2 \cite{Drees:1998pw}.
Thus any signal for pair--produced charged Higgs
bosons at LEP would be evidence
{\it against} the MSSM.

In the NMSSM where a Higgs singlet field
$N$ is added to the superpotential,
the above relation is modified to
\begin{equation}
M_{H^\pm}^2=M_A^2+M_W^2-\lambda_{_N} v^2.
\end{equation}
Here the $\lambda_{_N}$ contribution arises
from the $\lambda_{_N} NH_1H_2$
term in the superpotential. $M_A$ is now an entry in the extended
$3\times 3$ pseudoscalar mass matrix, and does not necessarily
correspond to the mass of a physical Higgs boson.
Clearly $M_{H^\pm} \le M_W$ is possible if
$\lambda_{_N}$ is suitably large.
Requiring that $\lambda_{_N}$ remains
perturbative up to the GUT scale,
Ref.~\cite{Drees:1998pw} showed that
$M_{H^\pm}\le M_W$ is possible for $1.7 \le \tan\beta\le 3.5$.
If $H^\pm$ is lighter than the $W$ boson,
its main decay modes are into $cs$ and $\tau\nu$.
Since $\Gm(H^- \to \tau^- \bar{\nu}_\tau)
/\Gm(H^- \to \bar{c}s)
\approx \tan^4 \beta (m_\tau^2/3 m_c^2)$,
the permitted region for $\tan\beta$
would give BR$(H^\pm \to \tau\nu_{\tau})\ge 90\%$, in clear
disagreement with the L3 signal.

Charged Higgs bosons also arise in
non--supersymmetric models with
two or more Higgs doublets. The Higgs potential
of such models is not restricted
by the constraints of supersymmetry
and thus there are no mass relations among the Higgs bosons.
In principle $M_{H^\pm}$ is a free parameter,
which may be chosen such that $M_{H^\pm}\le M_W$.
However, the observed decay rate of $b\to s \gm$
is known to provide strong constraints
on such a light $H^{\pm}$ (see below).

In a 2HDM with natural flavor
conservation \cite{Glashow:1977nt}
there are four distinct models depending
on how the Higgs doublets are coupled to the fermions
(the Yukawa couplings) \cite{Barger:1990fj}.
In Table I,
we summarize which type of fermions couple to
$H_1$ and $H_2$.
The Higgs sector of the
MSSM requires Model~II type couplings.
The Yukawa interaction for $H^+$ is given by
\begin{equation}
{\cal L} =\frac{g}{\sqrt{2}}
\left\{
\left( \frac{m_{d_i}}{M_W}
\right) X \bar{u}_{Lj} V_{ji} d_{Ri}
+
\left( \frac{m_{u_i}}{M_W}
\right) Y \bar{u}_{Ri} V_{ij} d_{Lj}
+
\left( \frac{m_{l_i}}{M_W}
\right) Z \bar{\nu}_{Li} e_{Ri}
\right\}
H^+ + H.c.\,.
\end{equation}
Here $u_L$ and $u_R$ ($d_L$ and $d_R$) respectively
denote left- and right-handed up (down)
type quark fields, $\nu_L$ is the left--handed neutrino field,
and $e_R$ the right--handed charged lepton field.
The $V$ is the CKM matrix.
Table II shows the couplings $X$, $Y$ and $Z$
in the 2HDM \cite{Barger:1990fj}.

In a MHDM with $N$ doublets ($N\ge 3$),
the couplings $X$, $Y$ and $Z$ are
arbitrary complex numbers which originate from
the mixing matrix for the charged scalar sector \cite{Albright:1980yc}.
In a model with $N$ doublets there are ($N-1$) $H^+$'s,
each with fermionic couplings $X_i$, $Y_i$ and $Z_i$
($i=1,2, \cdots,N-1$).
These couplings obey various sum rules 
due to the unitarity of the matrix which diagonalizes the
charged scalar mass matrix \cite{Grossman:1994jb}.
We shall only be concerned with
the lightest $H^{\pm}$, and thus drop the subscript $i$.

The phenomenology of charged Higgs bosons
differs from model to model.
Of particular importance is the $H^\pm$ contribution
to the decay $b\to s\gamma$ \cite{Bor,Hewett:1994bd}.
To leading order, its decay rate
is known to be
\beq
\Gm(b\to s \gm)
=\frac{\alpha_{em} G_F^2 m_b^5}{32\pi^4}
\left|
V^*_{ts} V_{tb}
\right|^2
\left|
\bar{D} (m_b)
\right|^2,
\eeq
where the $\bar{D}$ is the effective Wilson coefficient.
The $H^\pm$ contributions modify
the $\bar{D}$ into
\beq
\label{Dbar}
\bar{D}(M_W) =
\bar{D}_{SM} \left( \frac{m_t^2}{M_W^2}\right)
+ |Y|^2 \bar{D}_{YY}\left( \frac{m_t^2}{M_{H^\pm}^2}\right)
+ (XY^*) \bar{D}_{XY}\left( \frac{m_t^2}{M_{H^\pm}^2}\right)
\,.
\eeq
The analytic form of the functions $\bar{D}_{SM}$,
$\bar{D}_{YY}$, and $\bar{D}_{XY}$
at next to leading order in QCD
can be found in Ref.~\cite{Bor}.
In type II and II$'$ 2HDM's,
the value of $XY^*$ is fixed to be one,
which imposes a lower limit of
$M_{H^\pm} \ge 160$\,GeV,
with the bound becoming stronger
with increasing $|Y|$ \cite{Bor}.
In type I and I$'$ 2HDM's and a MHDM,
the absence of such a constraint permits a $H^\pm$
to be light enough to explain the L3 data.

Assuming $M_{H^\pm}=68$\,GeV, we now check whether
any of these models can accommodate
the branching ratios of the $H^\pm$
as observed by L3.
We define the following ratio:
\begin{equation}
\label{R}
\R \equiv
{\Gamma(H^\pm\to cs)\over \Gamma(H^\pm\to \tau\nu_{\tau})}
\approx {3|V_{cs}|^2(m_c^2|Y|^2+m_s^2|X|^2)
\over m_{\tau}^2|Z|^2},
\end{equation}
which is constrained by the L3 data to be $\R\approx 9$.
In the type I and I$'$ 2HDM's where $X=Y=\cot\beta$,
the term proportional to
$m_s^2$ may be neglected.
In type I 2HDM, $\R\approx 9$ cannot be attained
since the $\cot\beta$-dependence
cancels out in $\R$, leaving $\R\approx 0.5$.
In Model I$'$ one has $\R\le 0.5$ for $\tan\beta\ge 1$.
Although lower values of $\tan\beta$
would increase $\R$
($\R \approx 9$ is possible for $\tan\beta=0.45$),
such low values of $\tan\beta$
together with $M_H=68$\,GeV enhances too much the $H^\pm$
contributions to the $b\to s\gamma$ decay
(see Eq.~(\ref{Dbar})
and Table II) \cite{Bor}.
Thus we conclude
that neither type I nor I$'$ 2HDM can achieve $\R\approx 9$ as
required by the L3 data.

Higgs Triplet Models (HTM) are another source of $H^\pm$
\cite{Huitu:2001ut,Godbole:1995np}, and
have the added bonus of providing a mass for neutrinos
\cite{Mohapatra:1980ia}.
A $H^\pm$ composed dominantly of scalar triplet fields
only couples very weakly to quarks, rendering its
contribution to $b\to s\gamma$ negligible, and thus may be light.
However, such models usually predict enhanced BRs to leptons through
new Yukawa couplings \cite{Huitu:2001ut}, or exotic decays
$H^\pm\to W^{*}Z^{*}\to ffff$ \cite{Godbole:1995np}.
Hence $R\approx 9$ seems unlikely in a HTM.

A MHDM can easily obtain $\R\approx 9$,
provided $|Y|\approx 5|Z|$.
Since $|Y|$ and $|Z|$ are essentially free parameters,
one may choose $|Y|$ and $|Z|$ appropriately,
while simultaneously satisfying the constraints from
$b \to s \gm$ and $Z\to b\bar{b}$.
If $|X|$ is much larger than
$|Y|$ and $|Z|$ then $H^{\pm}\to cb$ becomes the
dominant channel since the CKM suppression of
$V_{cb}$ is well compensated
by the large ratio of $m_b/m_s$ \cite{Grossman:1994jb,Akeroyd:1995ga}.
We point out here that
DELPHI searches for events consistent with $H^\pm\to cs$
by imposing an anti--$b$ quark tag.
Note the possibility that a $H^\pm$ with a large
BR$(H^\pm\to cb$) might escape the DELPHI search strategy.

\section{Stau LSP interpretation of L3 excess}
\label{stau-L3}

In this section we show that the L3 excess
can be naturally explained by a LSP $\stau_1$.
Attributing the L3 excess to stau pair--production
is attractive in the sense that it would be a SUSY
explanation of the data, and a signal of a model which
generates a mass for neutrinos
(i.e., a $R$-parity violating model).
Unlike the Higgs case,
the $R$-parity violating Yukawa couplings
are not proportional to the fermion mass,
and thus the decay channels to light quarks (e.g.,
$\tilde{\tau}^- \to \bar{u} d$)
can be sizable. Note that the L3 conclusion of 90\% BR$(H^\pm \to c s)$
is based on the assumption of charged Higgs bosons.
In fact, the search is sensitive to any light quark,
e.g., $u \, d$.
The relatively large ratio of the hadronic to leptonic
BRs may be partially explained by the availability of
several unsuppressed hadronic decay channels.
As mentioned in Sec.~\ref{stau}, a LSP $\stau_1$ in a purely bilinear $R$-parity violating model
would also give $H^\pm$ like signals since it would decay via its
$H^\pm$ component. However, whether or not the observed BRs
could be obtained lies outside the present study, and would require a
careful analysis of the correlation between the $\stau-H^\pm$
mixing and the bilinear $R$-parity violating
parameters, the latter being strongly constrained from the
observation of neutrino oscillations.

The stau interpretation of the L3 excess
requires three conditions.
Firstly, the stau should be the LSP,
which is cosmologically permitted
in $R$-parity violating supersymmetric models.
In the R parity conserving MSSM 
the lightest neutralino $(\tchi_1^0)$ is the LSP and 
the stau would dominantly decay into
$\tau \tchi_1^0$. This gives rise to a signature incompatible
with the L3 data and the lower limit 
$M_{\stau} \gsim 80$~GeV has been obtained \cite{Groom:2000in}.
If the sneutrino is the LSP,
the strong mass constraint ($m_{\tilde{\nu}}\gsim 1$ TeV)
from the direct relic searches
in underground low-background experiments
\cite{Klapdor-Kleingrothaus:1998ya},
rules out the presence of such a light stau.
Secondly, in order to explain the observed decays into $qq'$,
$\stau_1$ must contain some $\stau_L$, since $\stau_R$
only decay to leptons (see Eq.~(\ref{Lg})).
We will see below that $\stau_1$ should be dominantly composed of
$\stau_L$ in order to comply with the observed cross--section.
Thirdly, both $\lm$ and $\lm'$ Yukawa couplings should
be non-zero in order to allow both hadronic and leptonic decay modes.
This is different from the widely applied assumption
that one $R$-parity violating operator is dominant at a time.
Present searches at LEP for $R$-parity violating
decays of scalar fermions
are also based on this one-at-a-time assumption. The current search
for a LSP stau only considers direct decays to
$qq'$ via a $\lambda'$ coupling, or direct decays
to $l\nu_i$ via the $\lambda$ coupling \cite{Abbiendi:2001is}.

Now let us check whether $\sigma(e^+e^-\to \stau_1^+\stau_1^-)$
can be compatible with that of $\sigma(e^+e^-\to H^+H^-)$.
For simplicity, we assume that the $\lm_{i33}$ coupling is
dominant over other $\lm$ couplings. Otherwise there would be extra
$t$--channel contributions.
Therefore the dominant contributions
are from $\gm$ and $Z$ gauge bosons in the $s$ channel,
which is the same as in the $R$-parity conserving MSSM.
In the absence of left-right mixing in the stau mass matrix,
and assuming that $\stau_1=\stau_L$,
the couplings $Z\stau_1^+\stau^-_1$ and
$\gamma\stau_1^+\stau^-_1$
are equal to the analogous couplings for $H^\pm$.
Therefore the cross--sections are the same if $\tilde{\tau}_1$
and $H^\pm$ have the same mass.
If $\stau_1$ has a component of $\stau_R$ then
$\sigma(e^+e^-\to \stau^+_1\stau_1^-)$
is reduced compared to
that for $\sigma(e^+e^-\to H^+H^-)$
since the $\stau^*_1 \stau_1 Z$ coupling
is proportional to
$ (\cos^2\theta_\stau-2 \sin^2\theta_W)$,
with $\theta_\stau$ being the left-right stau mixing angle.

To obtain $M_{\stau_1}=68$ GeV one merely requires the
various SUSY parameters to be chosen appropriately.
In the case of no mixing $(\theta_\stau=0)$,
requiring $M_{\stau_1}=68$ GeV limits
$m_{\tilde{\tau}_{_{L}}} \approx 53$,
49, and 48\,GeV
for $\tan\beta=3$, 10, and 50 respectively.
Including stau left-right mixing increases
the allowed region for $m^2_{\tilde{\tau}_{_{L}}}$.
Since the stau production cross--section
should not be decreased too much by the mixing,
we constrain the ratio of
$\sg ( e^+ e^- \to \stau_1^+ \stau^-_1)/
\sg ( e^+ e^- \to H^+ H^-)$ to be larger than $0.9$.
With $A_\tau=100$\,GeV and $\mu=100~ (200)$\,GeV,
Fig.~1 (2) exhibits the allowed region
in the $(m_{\tilde{\tau}_{_{R}}},m_{\tilde{\tau}_{_{L}}})$
plane after requiring $M_{\stau_1} = 68 \pm 2$\,GeV.
It can be easily seen that larger stau mixing,
which occurs for large
$\tan\beta$ and large $|\mu|$,
increases the allowed $m_{\tilde{\tau}_{_{L}}}$.

Finally we show that the LSP stau interpretation can
naturally explain the branching ratios of the
charged scalar as observed by the L3.
For the leptonic decay,
the stau can decay into $\tau\nu_e$ and $\tau\nu_\mu$
with non--zero $\lm_{i33}(i \neq 3)$.
Since $\lm_{133} (< 0.004)$ is rather strictly
constrained from the
bound on the mass of $\nu_e$ \cite{Groom:2000in,Hall:1984id}, we assume
that the main leptonic stau decay mode is into
$\tau \nu_\mu$.
For the hadronic stau decays,
we assume that all $\lm'_{3ij}$ are the same order of magnitude.
Then various decay channels are open;
$\stau_1\to ud, us, ub, cd, cs, cb$.
The ratio $\R$ defined in Eq.~(\ref{R}) becomes
\beq
\R \simeq
\frac{\sum_{i=1}^2\sum_{j=1}^3 |\lm'_{3ij}|^2}
{|\lm_{233}|^2}\,.
\eeq
The bounds on the $\lm_{ijk}$ and $\lm'_{ijk}$
have been obtained from various physical processes:
$\lm_{233} < 0.06$ is obtained from
$\Gm(\tau \to e \nu \bar{\nu})/
\Gm(\tau \to \mu \nu \bar{\nu})$ \cite{Groom:2000in,Barger:1989rk};
$\lm'_{311} < 0.16$ from
BR$(\tau\to\pi \nu_\tau)$ \cite{Bhattacharyya:1997nj};
$\lm'_{322} < 0.20$ from
$D^0-\bar{D}^0$ mixing \cite{Gupta:1997yt,Agashe:1996qm}.
Thus $\R \approx 9$ can be naturally accommodated
in the scenario of a LSP $\stau_1$.

\section{Distinguishing between $H^\pm$
and LSP $\tilde \tau_1$
at future colliders}
\label{diff}

In this section we will discuss how $H^\pm$ of the MSSM and a LSP
$\tilde \tau_1$ may be distinguished at future colliders. Note that
$H^\pm$ of the $R$-parity violating model under
consideration is expected to
possess a phenomenology very similar to that of the MSSM $H^\pm$, and so
our comments will be valid for both cases.
As discussed previously, a LSP $\stau_1$ has more
possibilities for $ff'$ decays, which are proportional to
arbitrary couplings $\lambda$ and $\lambda'$.
In general there
would be no tendency to decay into the heaviest allowed fermion,
unlike the case for $H^\pm$.
For a given $M_{H^\pm}$ the BRs of $H^\pm$
are calculable functions of $\tan\beta$ and hence are much
more predictable, with $H^\pm\to tb$
dominating for $M_{H^\pm}\ge m_t+m_b$
and $H^\pm\to \tau\nu_{\tau}$ dominating for $M_{H^\pm}\le m_t+m_b$.
The $H^\pm\to cs$ decays can compete with $H^\pm\to \tau\nu_{\tau}$ only
for very low $\tan\beta$,
which is already disfavored experimentally.
Therefore sizeable BRs for
$\stau_1\to e\nu_i,\mu\nu_i$ would
be clear signals for $\stau_1$,
as would enhanced BRs to light quarks
$\stau_1\to ud,cs$ etc. Note that these latter
decays may also dominate for the region $m_{\stau_1}\ge m_t$, while
for $m_{H^\pm}$ the dominate decay would be $H^{\pm}\to tb$, which
gives a very different signature.
A high--energy $e^+e^-$ collider would be
an ideal place to distinguish the flavor of the jets from
$\stau_1\to ff'$ decays.

As pointed out in the previous section,
if $\stau_1$ is mainly $\stau_R$
then its production cross--section $\sigma(e^+e^-\to \stau^+_R\stau^-_R)$
would be suppressed compared to that for
$\sigma(e^+e^-\to H^+H^-)$.
In general, one would expect
$\stau_1$ to be a mixture of $\stau_R$ and $\stau_L$,
and so there would always be some suppression
compared to the $H^+H^-$ production.
Given the expected high luminosity of
proposed linear colliders, even relatively small differences
in the rates might be observable. However,
one-loop corrections to $\sigma(e^+e^-\to H^+H^-)$ should not be
ignored since these can be up to $30\%$ \cite{Arhrib:1999gr}, thus
rendering difficult this method of distinguishing $\stau_1$ and
$H^{\pm}$.

At hadron colliders, such as the Tevatron and LHC,
a sufficiently light $H^\pm$ may be produced by the
decay $t\to H^\pm b$. At the Tevatron Run II discovery in this channel
is possible for small or large
$\tan\beta$ \cite{Carena:2000yx}, with improved coverage at the LHC.
Therefore a $H^\pm$ signal would provide information
on $\tan\beta$. However, in the LSP $\stau_1$ scenario,
the decay $t\to \stau_1b$ may be open with a rate depending
on the arbitrary coupling $\lambda'_{333}$
\cite{Agashe:1996qm,Erler:1997ww}. Hence
$H^\pm$ like signals
and corresponding measurements of $\tan\beta$ in this channel
should be interpreted with care.

At the LHC the discovery of $H^{\pm}$ for
$M_{H^\pm}\ge m_t$ is considered challenging\cite{Roy:2001mj}.
Currently the most
effective method is to use the production mechanism
$gg(q\overline q) \to H^\pm tb$ followed by $H^\pm\to \tau\nu_{\tau}$ decay
\cite{Roy:1999xw}.
This method offers reasonable detection prospects
for $\tan\beta\ge 15$, where
BR($H^\pm\to \tau\nu)\approx 10\%$ for this region.
Using the above production mechanism followed by the
decay $H^\pm \to tb$ requires highly efficient $b$ tagging
\cite{Borzumati:1999th} due to the huge hadronic backgrounds.
For $\stau_1$, the analogous mechanism
$gg(q\overline q) \to \stau_1tb$ can be used \cite{Borzumati:1999th},
and offers sizeable cross--sections for $\lambda'_{333}\ge 0.01$.
Detection of a LSP $\stau_1$
in its light hadronic decay modes would be unlikely
due to the large QCD background
but $\stau_1$ decay to $l\nu_i$ should provide a very promising
signature. For $l=e,\mu$ the signature would be distinct
from that of $H^\pm$.
For $l=\tau$ there is the possibility of a much larger BR$(\stau_1\to
\tau\nu_i$), which would enhance the signal size compared to that
for $H^\pm$. Note also that $\stau_1$ may be produced as a
$s$-channel resonance at hadron colliders \cite{Dimopoulos:1990fr}, while
the corresponding rates for $H^\pm$ would be very small due to the
suppressed Yukawa couplings to the light quarks.

Finally we mention the possibility of very different
lifetimes for $H^{\pm}$ and $\stau_1$.
In general $H^\pm$ is expected to decay promptly,
especially if $H^\pm\to tb$ decays are open.
Since the $\stau_1$ decay rates are proportional to the
arbitrary $\lambda$ and $\lambda'$ couplings, the various
partial widths may be very suppressed. The LEP searches, assuming one
coupling is dominant, are sensitive to
$\lambda,\lambda' \gsim 10^{-5}$
\cite{Abbiendi:2001is}. If $\stau_1$ possessed very similar BRs to
$H^\pm$, (e.g, $\stau_1\to tb$ dominating for
$M_{\stau_1}\ge m_t$), the lifetimes
would be very different if $\lambda'_{333}$ were
considerably less than the corresponding $H^\pm tb$ Yukawa
coupling. This might leave an observable decay length in the detector,
which could not be attributed to a $H^\pm$.
If $\lambda,\lambda' \lsim 10^{-5}$ then $\stau_1$ would decay outside the
detector, but could be detected as a long lived charged particle
\cite{Ackerstaff:1998si}.

\section{Conclusions}
\label{conclusion}

In the context of a $R$-parity violating model we have
studied the phenomenological implications of
the assumption that the lightest supersymmetric particle (LSP)
is the lightest stau ($\stau_1$). In such a model the LSP is unstable and
is not in conflict with the usual cosmological requirement
that any LSP should be charge- and color-neutral, conditions
which apply only to stable particles.

The stau LSP assumption has important implications
for both $H^\pm$ and stau searches.
A left--handed stau possesses many phenomenological
similarities with $H^\pm$.
Two major differences between $\stau_1$ and $H^\pm$
are the mass spectrum and the decay modes.
Whereas the presence of a light
charged Higgs boson ($M_{H^\pm} < M_W$) in
many popular extensions of the SM
is severely constrained by the supersymmetric structure of the
theory and/or other phenomenological constraints such as
the decay $b \to s \gm$, a light stau is naturally accommodated
within a reasonable SUSY parameter space. In models
with anomaly-mediated breaking of supersymmetry the stau
would be a natural candidate for the LSP,
provided that $R$-parity
is also violated.
It is feasible that a LSP stau is lighter than the charged Higgs boson,
and thus may be observed first at future colliders.
The distinction of $\stau_1$ signals from $H^\pm$ signals
is, in principle, possible by examining the decay modes.
For a given $H^\pm$ mass, its decays are essentially determined by
the decaying fermion mass and the parameter $\tan\beta$.
In contrast, $\stau_1$ decays possess many more possibilities
due to the arbitrariness of $R$-parity violating
couplings $\lm$ and $\lm'$.

One of the most remarkable implications in our scenario
is the possibility that a LSP stau may imitate $H^\pm$.
In particular, when both $R$-parity violating
couplings, $\lm$ and $\lm'$, are non-zero,
a light LSP stau may possess $H^\pm$ like hadronic and leptonic BRs
and thus may be misconceived as the charged Higgs boson.
This misidentification possibility has received little attention
due to the usual simplifying assumption
that one $R$-parity violating operator is dominant at a time.
One possible example of $H^\pm$ misidentification
is the recently reported L3 excess of $4.4\sigma$ in the
search for pair-produced charged Higgs bosons.
Attributing the signal to $H^\pm$ production
has severe problems in many popular models such as the MSSM,
NMSSM, and 2HDM.
We have shown that the LSP stau interpretation offers a more
attractive explanation of the data and simply requires
that the LSP $\stau_1$ is mainly left--handed
with simultaneous non-zero values for the couplings $\lm$ and $\lm'$.
The L3 data can be summarized by the following three characteristics:
pair-produced singly charged particles with mass around
68\,GeV; a production cross-section comparable to that for $H^\pm$;
decay BRs of $90\%$ into light quarks and
$10\%$ into a tau lepton with a neutrino.
All three features can be explained in the LSP $\stau_1$
scenario without any fine-tuning, and within the
experimental bounds on $R$-parity violating couplings
and SUSY breaking scalar masses.

Finally, we have discussed how to distinguish
a stau LSP from a $H^\pm$ signal at future colliders.
Firstly, anomalous decay modes into
light fermions (e.g., $ud$, $e\nu$, or $\mu\nu$)
would be a robust signal of the LSP stau.
Such decays are permitted since the $R$-parity violating couplings
are in principle independent of the fermion mass, in contrast
to the case of
$H^\pm$ which has a tendency to decay into the heaviest
available fermions. Secondly,
the tree--level pair-production cross--sections for the LSP stau
and $H^\pm$ at future $e^+e^-$ collider may differ
since the mixing between the left- and right-handed
stau's decreases the $\stau_1$ cross--section compared to that
for $H^\pm$. However, the one--loop corrections to these rates
can be sizeable and complicate this method of distinguishing
$\stau_1$ from $H^\pm$. Note also that $\stau_1$ may be produced as a
$s$-channel resonance at hadron colliders \cite{Dimopoulos:1990fr}, while
the corresponding rates for $H^\pm$ would be very small due to the
suppressed Yukawa couplings to the light quarks.

Thirdly, the lifetime of $\stau_1$ may be
much longer than that for $H^\pm$. This could
leave an observable decay length in the detector
which could not be attributed to $H^\pm$.
We stress the fact that a LSP $\stau_1$ may mimic the phenomenology
of $H^\pm$ and thus signals for $H^\pm$ and corresponding measurements
of the Higgs sector parameters (e.g., $\tan\beta$) should be
interpreted with care.

\acknowledgements
We thank T. Han, M.A. Diaz and F.M. Borzumati for useful comments.
C.L. is supported in part by the National Natural Science Foundation of China 
with grant no. 10047005.

\begin{table}[htb]
\centering
\begin{tabular} {|c|c|c|c|c|}
 & Model~I & Model~I$'$ & Model~II & Model~II$'$  \\ \hline
u (up--type quarks)   & 2 & 2 & 2 & 2 \\ \hline
d (down--type quarks) & 2 & 2 & 1 & 1 \\ \hline
e (charged leptons)   & 2 & 1 & 1 & 2 \\
\end{tabular}
\vskip .5cm
\caption{The four distinct structures of the 2HDM.}
\vskip 4.5cm
\centering
\smallskip
\begin{tabular} {|c|c|c|c|c|}
 & Model I & Model I$'$ & Model II & Model II$'$  \\ \hline
$X$ & $-\cot\beta$ & $-\cot\beta$ & $\tan\beta$ & $\tan\beta$ \\ \hline
$Y$ & $\cot\beta$  & $\cot\beta$ & $\cot\beta$ & $\cot\beta$ \\ \hline
$Z$ &  $-\cot\beta$  & $\tan\beta$  & $\tan\beta$
& $-\cot\beta$ \\
\end{tabular}
\vskip .5cm
\caption{The values of $X$, $Y$ and $Z$ in the 2HDM.}
\end{table}

\vskip 2cm
\begin{center}
\begin{figure}[htb]
\hbox to\textwidth{\hss\epsfig{file=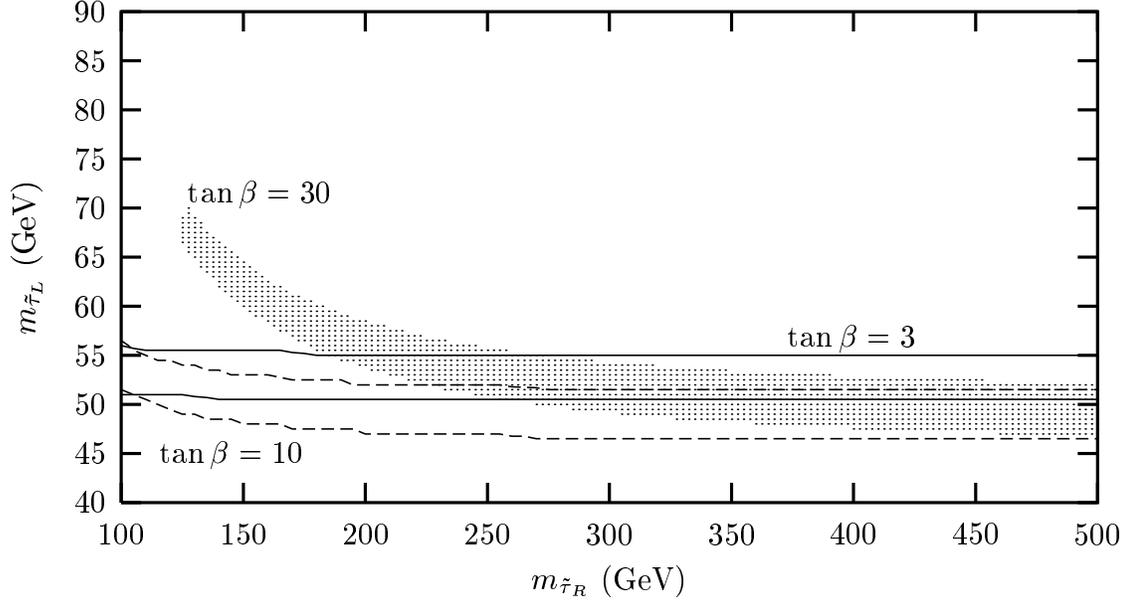,height=8cm}\hss}
\smallskip
\caption{ Allowed region
in the $(m_{\tilde{\tau}_{_{L}}},m_{\tilde{\tau}_{_{R}}})$
plane by the mass constraint $M_\stau = 68 \pm 2$~GeV
and $\sg ( e^= e^- \to \stau^+ \stau^-)/
\sg ( e^= e^- \to H^+ H^-)> 0.9$.
$A_\tau=\mu= 100$~GeV.
The solid-lined-band is for $\tan\beta=3$,
the dashed-lined-band for $\tan\beta=10$,
and the dotted band for $\tan\beta=30$.
}
\label{fig1}
\end{figure}
\end{center}

\begin{center}
\begin{figure}[htb]
\hbox to\textwidth{\hss\epsfig{file=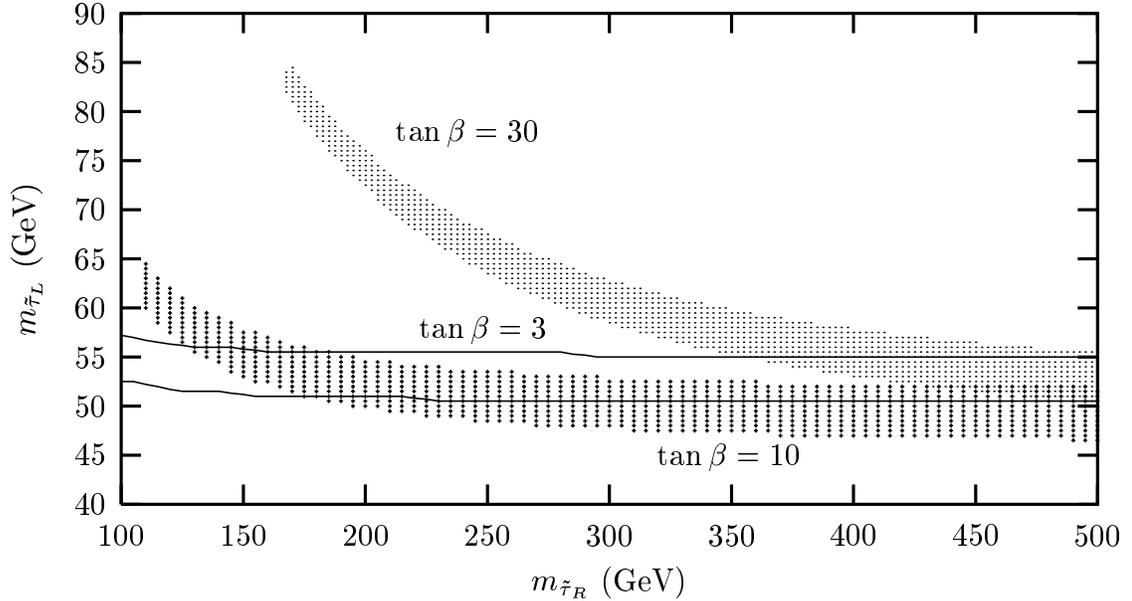,height=8cm}\hss}
\smallskip
\caption{ The same plot for $\mu=200$~GeV.
The band with larger dots is for $\tan\beta=10$.
}
\label{fig2}
\end{figure}
\end{center}
\end{document}